\documentclass{WileyMSP-template}

\usepackage[comma,numbers,square,sort&compress]{natbib}
\usepackage{epstopdf}
\usepackage{graphicx} 
\usepackage{dsfont}
\usepackage{mathrsfs}
\usepackage{amsmath} 
\usepackage{amssymb}  
\usepackage{amsfonts}
\usepackage{xcolor}
\usepackage{epstopdf}
\usepackage{subfigure}
\usepackage{amsthm}
\usepackage{amsmath,amssymb,amsfonts}
\usepackage{algorithmic}
\usepackage{textcomp}
\usepackage{flushend}
\usepackage{ragged2e}

\newtheorem{remark}{Remark}

\begin{document}

\pagestyle{fancy}
\rhead{}

\title{Cooperation between coherent control and noises\\ in quantum metrology}

\maketitle


\author{Yu~Chen, Zibo~Miao* and Haidong~Yuan*}



\begin{affiliations}
Y. Chen\\
Mechanical and Automation Engineering, The Chinese University of Hong Kong\\
Tencent Lightspeed Quantum Studios, Tencent, Shenzhen, 518057, China\\

Z. Miao\\
School of Mechanical Engineering and Automation, \\
Harbin Institute of Technology, Shenzhen, 518055, China\\
E-mail: miaozibo@hit.edu.cn

H. Yuan\\
Mechanical and Automation Engineering, The Chinese University of Hong Kong\\
E-mail: hdyuan@mae.cuhk.edu.hk

\end{affiliations}


\keywords{quantum metrology, cooperative scheme, improved precision limit}

\begin{abstract}

 \justifying \noindent In this paper we study the cooperation between coherent control and noises in open spin systems, aiming to demonstrate that such cooperation can provide new possibilities for parametrization in quantum metrology. The cooperative scheme proposed here outperforms the standard scheme, with higher precision achieved. More specifically, we illustrate the effect of cooperative interaction between coherent control and noises in conventional single-spin systems described by Lindblad master equations, with the magnitude of a magnetic field taken as the parameter to be estimated and encoded in the noisy dynamics besides the Hamiltonian. The scenarios of both spontaneous emission and dephasing noise have been analyzed, where the associated quantum Fisher information has been given. Furthermore, it has been demonstrated that in the realm where quantum metrology is mostly applied in practice, the precision limit under the cooperative scheme in the presence of noises can surpass the ultimate precision limit under the unitary dynamics. On the other hand, multiple-spin systems have also been considered. We show that the coupling between different spins can help realize non-local parametrization under the cooperative scheme, with the ground state made entangled. This thus leads to the improvement of  precision limit, which has been proved and visualized in our paper.
\end{abstract}


\section{Introduction}
\justifying To study the precision limit of measurement and estimation in science and technology, metrology has been playing a central role. In recent years, quantum metrology, which exploits quantum mechanical effects to achieve better precision in comparison to purely classical schemes, has attracted increasing attention and has seen wide applications in a variety of research fields \cite{giovannetti2011advances,giovannetti2006quantum,anisimov2010quantum,braunstein1996generalized,escher2012general,demkowicz2014usin,demkowicz2012elusive,
	schnabel2010quantum,ligo2011gravitational,joo2011quantum,higgins2007entanglement,lugiato2002quantum,morris2015imaging,roga2016security,tsang2016quantum,buvzek1999optimal,	leibfried2004toward,roos2007designer,ludlow2015optical,borregaard2013near,shapiro2009quantum,lopaeva2013experimental,dowling1998correlated,JCJD2021,huelga1997improvement,chin2012quantum,HallPRX,Berry2015,Alipour2014,Beau2017}, such as quantum phase estimation \cite{escher2012general,joo2011quantum,anisimov2010quantum,higgins2007entanglement}, gravitational wave detection \cite{schnabel2010quantum,ligo2011gravitational}, quantum imaging \cite{lugiato2002quantum,morris2015imaging,roga2016security,tsang2016quantum}, atomic clock synchronization \cite{buvzek1999optimal,leibfried2004toward,roos2007designer,ludlow2015optical,borregaard2013near}, quantum target detection \cite{shapiro2009quantum,lopaeva2013experimental}, and quantum gyroscopes \cite{dowling1998correlated,JCJD2021}.

\begin{figure}[htb]
	\centering\includegraphics[width=0.5\textwidth]{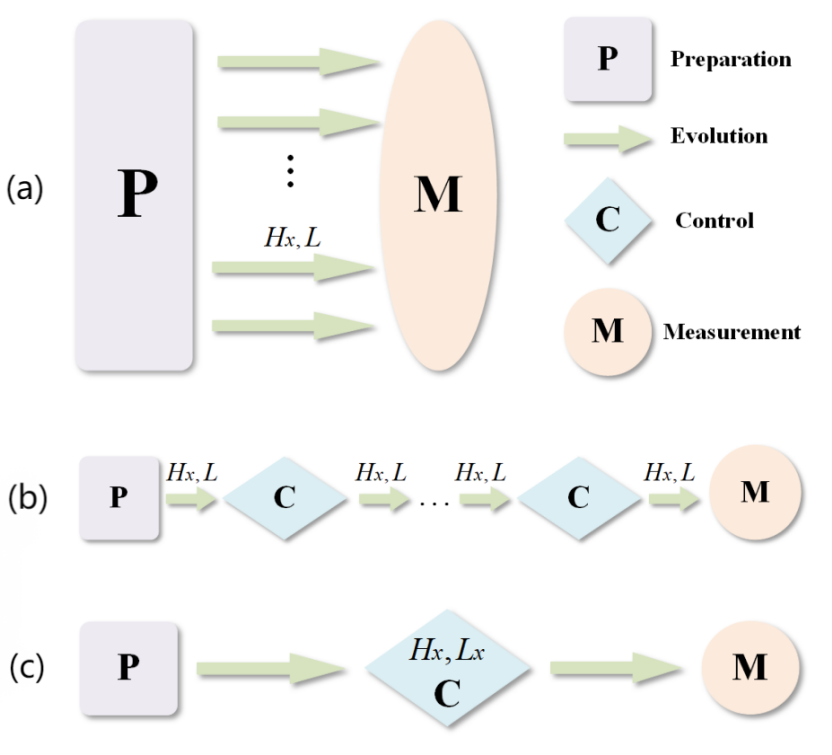}
	\caption{(a) parallel scheme, (b) controlled sequential scheme, and (c) cooperative scheme.  The three schemes are equivalent under unitary dynamics ($x$ denotes the parameter to be estimated). Specifically, in the cooperative scheme $H_x$ and $L_x$ denote the Hamiltonian and noisy operator with $x$ encoded by means of coherent control.}
	\label{fig:scheme}
\end{figure}

Standard schemes of quantum metrology, as shown in Fig. \ref{fig:scheme}(a) and Fig. \ref{fig:scheme}(b), can be explained by the canonical example where spins are used to estimate the magnitude of a magnetic field. Without noises, the dynamics of each spin in the magnetic field is governed by the Hamiltonian $H=B_z\sigma_z$ ($\sigma_x$, $\sigma_y$ and $\sigma_z$ are the Pauli operators that can generate the special unitary group $SU(2)$ together with the identity operator $I$), with $B_z$ the parameter to be estimated (here the gyromagnetic ratio of the spin has been absorbed in the magnitude $B_z$). If $N$ spins are prepared in the Greenberger–Horne–Zeilinger state (GHZ state), $\frac{1}{\sqrt{2}}(|00\cdots0\rangle+|11\cdots1\rangle)$, and they evolve under the unitary dynamics for the time duration $t$, the final state is then $\frac{1}{\sqrt{2}}(e^{iNB_zt}|00\cdots 0\rangle+e^{-iNB_zt}|11\cdots 1\rangle)$ ($i$ denotes the imaginary unit), which achieves the quantum Fisher information (QFI) $F_Q=4N^2t^2$ \cite{Holevo, helstrom1976quantum}. According to the quantum Cram\'er-Rao bound \cite{Holevo, helstrom1976quantum}, the standard deviation for any unbiased estimation, $\delta \hat{B_z}=\sqrt{\mathbb{E}[(\hat{B_z}-B_z)^2]}$ ($\mathbb{E}[\cdot]$ denotes the expectation), is bounded below by the quantum Fisher information as $\delta \hat{B_z}\geq \frac{1}{\sqrt{mF_Q}}\geq\frac{1}{\sqrt{m}2Nt}$, with $m$ being the number of times that the procedure is repeated. This precision limit can be equivalently achieved by letting a single spin evolve under the unitary dynamics for the time duration $T=Nt$. The precision limit, $\frac{1}{\sqrt{m}2T}$, which is referred to as the Heisenberg limit, provides the minimum standard deviation for any unbiased estimation with a given time $T$. It is also known that  the Heisenberg limit, $\frac{1}{\sqrt{m}2T}$, can not be surpassed by merely adding extra terms to the Hamiltonian (i.e. the Hamiltonian is given by $H=B_z\sigma_z+H_C(t)$ where $H_C(t)$ is determined by externally added control signals) \cite{giovannetti2006quantum,boixo2007generalized,yuan2015optimal,yuan2016sequential}.  
On the other hand, it is worth mentioning that there is established work on the effect of adding control under noisy dynamics, where different control techniques, including quantum error correction \cite{Dur2014,Arrad2014,Kessler2014,Unden2016,Zhou2018,SMMN2021}, dynamical decoupling \cite{Schmitt832,Boss837,SekatskoNJP2016,LangPRX2015,Taylor2008,Cooper2014} and optimal control  \cite{LiuSingle, LiuMulti,LZCWY2022}, are explored, with the purpose of obtaining improvement in the precision limit.  The control signals are mainly employed to either eliminate or suppress the noises in order to recover the performance of controlled unitary evolution since in general noises are regarded to be harmful. It is thus not surprising that the performance of controlled system evolution in the presence of noises is worse than the performance of controlled unitary evolution without noises. 

By contrast, in this paper we show that there are benefits provided by the interplay between coherent control and noises that can contribute to parameter estimation. Instead of suppressing the noises, our schemes can actively make use of the noises, as depicted in Fig. \ref{fig:scheme}(c) where $H_x$ and $L_x$ indicate that both the Hamiltonian and noisy operator are encoded with the parameter to be estimated. Under such cooperative schemes, the precision limit can go beyond the limit under above-mentioned standard schemes in quantum metrology. More precisely, we show that although coherent control or noises alone may not be utilized separately to improve the precision limit according to a given system evolution, the cooperative interaction between them can make a significant contribution. In more concrete terms, through the interplay between coherent control and noises, the parameter to be estimated can be encoded into multiple components of the dynamics, differing from the standard scheme where the parameter is usually merely encoded in one part of the dynamics (either in the Hamiltonian or in the noisy operator). This opens many possibilities to improve the precision limit. By analyzing canonical systems in quantum metrology, we show that the precision limit under the cooperative scheme proposed in this paper can beat its counterpart in the standard scheme. Furthermore, in a certain region which will be specified in the following sections of this paper, the precision limit under the cooperative scheme can even surpass the highest precision limit under the ideal unitary dynamics. 

There have been relevant studies on environment-assisted quantum metrology \cite{Goldstein2011,Cappellaro2012}, where the environment is assumed to consist of many spins that can not be controlled individually (but may be controlled collectively). These spins can be parameterized by unknown fields as well \cite{Goldstein2011,Cappellaro2012}. By designing proper pulses, the information accumulated by the spins in the environment can be transferred back to the probe state for the improvement of estimation precision limit \cite{Goldstein2011,Cappellaro2012}. These studies are different from the cooperative scheme studied in this paper, where we do not assume the environment to be composed of spins that can participate in the parametrization process on their own. The cooperative scheme thus works under Markovian dynamics where there is no back flow of information from the environment to the system.

In Section \ref{sec:SSCS}, we present the canonical single-spin system model in quantum metrology that is firstly considered in this paper. Then both the scenario of spontaneous emission and the scenario of dephasing noise are taken into consideration. Detailed derivation of the corresponding quantum Fisher information under the cooperative scheme is provided, and thus the statement that the cooperative scheme can outperform the standard scheme is rigorously proved. The effect of cooperative interaction between coherent control and noises has also been visualized in comparison to the standard scheme as well as the Heisenberg limit. This is followed by Section \ref{sec:TSD} where the canonical two-qubit system model is considered. In contrast to the standard scheme, the coupling between different qubits can help shape the ground and excited states, making them more sensitive to the parameter to be estimated under the cooperative scheme. More concretely, the ground state can be made entangled due to the interplay between coherent control and noises, which generates non-local parametrization resulting in the improvement of precision. Finally, Section \ref{sec:Con} provides some concluding remarks and future research directions.

\section{Enhanced precision of parameter estimation for single-spin systems \\under the cooperative scheme}
\label{sec:SSCS}
In the absence of noises, by preparing a spin in the state $\frac{1}{\sqrt{2}}(|0\rangle+|1\rangle)$ and letting it evolve under the unitary dynamics governed by the Hamiltonian $H=B_z\sigma_z$ ($B_z$ is the parameter to be estimated) for the time duration $T$, one can obtain the final state $\frac{1}{\sqrt{2}}(e^{iB_zT}|0\rangle+e^{-iB_zT}|1\rangle)$ with the quantum Fisher information $F_Q=4T^2$ \cite{Holevo,helstrom1976quantum}.

However, in practice, noises are inevitable in open quantum systems whose dynamics can be described by the following Lindblad master equation
\begin{align}
	\label{eq:bothmaster}
	\dot{\rho} =  -i[H(B_z),\rho]+ \mathcal{L}(\rho) ,\
\end{align}
where
\begin{align*}
	\mathcal{L}(\rho) = \gamma\left(L\rho L^{\dagger}-\frac{1}{2}\left\{L^{\dagger} L,\rho\right\}\right)
\end{align*}
is the super-operator that describes the interaction between the system and environment.
The system Hamiltonian is denoted by $H(B_z)$, with $B_z$ being the parameter to be estimated in quantum metrology. $L$ is the noisy operator that defines the type of noises, and $\gamma$ denotes the corresponding decay rate ($[,\cdot,]$ and $\{,\cdot,\}$ in Eq.~\eqref{eq:bothmaster} denote the commutator and anti-commutator respectively). There are in general two types of noises, namely spontaneous emission and dephasing. We first consider the scenario of spontaneous emission. Since the physical parameters herein have SI units, the units are omitted in the following analysis.

\subsection{Spontaneous emission}
In the case of spontaneous emission, the dynamics of a spin can be described by the following master equation
\begin{eqnarray}
	\dot{\rho} = -i[H\left(B_z\right),\rho]+\gamma\left(\sigma_{-}\rho\sigma_{+}-\frac{1}{2}\left\{ \sigma_{+}\sigma_{-},\rho\right\} \right),
	\label{eq:masteq_spon}
\end{eqnarray}
where the Hamiltonian $H\left(B_z\right)=B_z\sigma_z$,  $\sigma_{+}=(\sigma_{x} + i\sigma_y)/2 = |1\rangle\langle 0|$ and $\sigma_{-}=(\sigma_{x} - i\sigma_y)/2 = |0\rangle\langle 1|$ are the raising and lowering operators in the basis  
$\{|0\rangle, |1\rangle\}$ respectively.

Under standard schemes, the maximal quantum Fisher information with regard to the magnitude of a magnetic field that can be achieved at the time $T$ is $F_Q=4e^{-\gamma T}T^2$ \cite{yuan2015quantum}, which is always smaller than the standard Heisenberg limit $4T^2$. 

The coherent part and noisy part of dynamics are usually independent from each other. And typically only one part of the dynamics contributes to the parametrization of the probe state under standard schemes. In fact, without coherent control, the lowering and raising operators, $\sigma_{-}=|0\rangle\langle 1|$ and $\sigma_{+}=|1\rangle\langle 0|$ in Eq. (\ref{eq:masteq_spon}), are independent from the magnitude of the magnetic field. The parameter $B_z$ to be estimated is then only encoded in the Hamiltonian. 

However, by adding proper coherent control, as we are going to show, the coherent part and the noisy part can contribute to the parametrization of the probe state cooperatively. 
If we add a controlled field along the $X$-direction, the Hamiltonian becomes
	\begin{equation}
		\label{eq:depncH}
		H(B_z) =B_z \sigma_z + B_x \sigma_x.
	\end{equation}
	Thus the ground and excited states associated with the Hamiltonian in Eq.~\eqref{eq:depncH} are $| g\rangle = -\sin{\frac{\theta}{2}}| 1 \rangle + \cos{\frac{\theta}{2}}| 0 \rangle$ and $| e \rangle = \cos{\frac{\theta}{2}}| 1 \rangle + \sin{\frac{\theta}{2}}| 0 \rangle$ respectively, with $\theta = \arctan \frac{B_x}{B_z}$. The unknown parameter $B_z$ is encoded in the new basis $| g\rangle$ and $|e\rangle$. The dynamics can then be written as
	\begin{equation}
		\label{eq:interplay}
		\dot{\rho} = - i[B_z \sigma_z + B_x \sigma_x, \rho] + \gamma(\sigma_{-}^H \rho \sigma_{+}^H -\frac{1}{2}\{\sigma_{+}^H\sigma_{-}^H, \rho\}),
	\end{equation}
	where $\sigma_{+}^H = | e\rangle \langle g |$, $\sigma_{-}^H = | g \rangle \langle e |$ are defined in the new basis of  $\{ |g\rangle, |e\rangle \}$, which contains the parameter.
	The interplay between coherent control and the noisy part thus encodes the parameter into multiple components of the dynamics, contributing to the parametrization of the probe state cooperatively to achieve a higher precision limit than that under the standard scheme.  Please note that such a system possesses a stead state $|g\rangle$ which gives the quantum Fisher information
	\begin{align}
		F_Q(| g\rangle) = 4(\langle \partial_{B_z} g |\partial_{B_z} g\rangle -|\langle \partial_{B_z} g| g \rangle|^2) = \frac{B_x^2}{(B_x^2+B_z^2)^2}.
		\label{eq:ground_single}
	\end{align}
	
	To calculate the precision limit that can be obtained under this cooperative scheme, we prepare the probe state at $\frac{|0\rangle+|1\rangle}{\sqrt{2}}$ in the original basis spanned by $|0\rangle$ and  $|1\rangle$. We then add the controlled field and let the spin evolve under the dynamics governed by Eq. (\ref{eq:interplay}) for the time duration $T$.
	Furthermore, in this case the quantum state at $T$ can be analytically obtained, which, in the new basis spanned by $| g \rangle$ and $e\rangle$, can be written as
	\begin{eqnarray}
		\rho_T = \left(\begin{array}{cc}
			\frac{1}{2}(1+\sin\theta)e^{-\gamma T} & \frac{1}{2}\cos\theta e^{-\frac{1}{2}(\gamma+4i \Delta)T} \\\frac{1}{2}\cos\theta e^{-\frac{1}{2}(\gamma - 4i \Delta)T} & 1-\frac{1}{2}(1+\sin\theta)e^{-\gamma T}
		\end{array}\right),
	\end{eqnarray}
	where $\Delta = \sqrt{B_z^2+B_x^2}$.
	
	This state can also be written in the original basis of $\{|0\rangle,|1\rangle\}$ through the following transformation
	\begin{eqnarray}
		\label{eq:transform_basis}
		\left(\begin{array}{c}
			| e\rangle \\
			| g\rangle
		\end{array}\right) = \left(\begin{array}{cc}
			\cos \frac{\theta}{2} & \sin \frac{\theta}{2} \\
			- \sin \frac{\theta}{2} & \cos \frac{\theta}{2}
		\end{array}\right) \left(\begin{array}{c}
			| 1\rangle \\
			| 0\rangle
		\end{array}\right).
	\end{eqnarray}
	
	The quantum Fisher information for $\rho_T$ can be obtained from the formula $F_Q = \mathrm{tr}[(\partial_{B_z} \rho_T)^2]+\frac{\mathrm{tr}[(\rho_T \partial_{B_z} \rho_T)^2]}{\mathrm{det}\left(\rho_T\right)}$ (see e.g. \cite{dittmann1999explicit}), which gives
	\begin{align}
		F_Q(\rho_{T}) &= \frac{1}{16 \Delta ^2}  e^{-\gamma T} \left[-24 \sin \theta +8 \sin 3 \theta \right. \nonumber\\
		&\left.+8 \cos 2 \theta  (4 \Delta ^2 T^2+1)+\cos 4 \theta  (8 \Delta ^2 T^2-1)\right.\nonumber\\ 
		&\left.+64 \Delta  T \sin \theta  \cos ^2\theta  e^{-\frac{1}{2} \gamma  T} \sin (2 \Delta  T) (\sin (\theta )\!-\!e^{\gamma  T}\!+\!1) \right.\nonumber \\ 
		&\left. +32 \sin^2\theta  e^{-\frac{1}{2} \gamma  T} \cos (2 \Delta  T) (\sin \theta  (e^{\gamma  T}-1)-1)\right.\nonumber\\
		&\left.-16 (\sin \theta +2) \sin ^3\theta  \sinh (\gamma  T) \right. \nonumber \\
		& \left. -8 \sin ^2\theta  (-4 \sin \theta +\cos 2 \theta -5) \cosh (\gamma  T)\right.\nonumber\\
		&\left.+2 \sin ^2(2 \theta ) \cos (4 \Delta  T)-7\right].
		\label{eq:Fqcomplicated}
	\end{align}
	
	Since the quantum Fisher information regarding the state $|g\rangle$ grows when the value of $B_z$ decreases, the improvement of precision under the cooperative scheme is thus more obvious in the region where $B_z$ is small. Moreover,  quantum metrology is mostly used for small values of $B_z$. And if $B_z$ is not small, one can always use adaptive control to shift $B_z$, by adding control fields along the opposite direction.
	
	Although the expression for $F_Q(\rho_T)$  in Eq.~\eqref{eq:Fqcomplicated} is a bit complicated, for a short period of time $T$, we can approximate $F_Q(\rho_T)$ with a low-order Taylor expansion to gain some insight. The first and second order terms of the low-order Taylor expansion can be obtained by calculating the first and second derivatives as follows
	\begin{align}
		\dot{F}_Q(0) &\equiv \frac{d}{dT}F_Q(\rho_{T})\big|_{T=0}\nonumber\\
		& = \frac{\gamma  \sin ^2(\theta ) \cos ^2(\theta )}{\Delta ^2} = \gamma \frac{B_x^2 B_z^2}{(B_x^2+B_z^2)^3},\\
		\ddot{F}_Q(0) &\equiv \frac{d^2}{dT^2}F_Q(\rho_{T})\big|_{T=0} \nonumber\\
		&= \frac{\gamma ^2 \left(16 \sin ^3(\theta )-10 \cos (2 \theta )+3 \cos (4 \theta )+7\right)}{8 \Delta ^2}+8 \nonumber\\
		&= \frac{\gamma^2\sin^2 \theta}{2 \Delta^2}(6 \sin^2 \theta + 4\sin \theta -1) + 8.
	\end{align}
	Therefore, the Taylor expansion of the quantum Fisher information $F_Q(\rho_{T})$ for a small period of time $T$ is then
	\begin{align}
		F_Q(\rho_{T}) & = \dot{F}_Q(0)T+\frac{1}{2}\ddot{F}_Q(0)T^2+o(T^3)\nonumber\\
		& = 4T^2 + \frac{\gamma  \sin ^2(\theta ) \cos ^2(\theta )}{\Delta ^2}T+ \frac{\gamma^2 \sin^2 \theta}{4 \Delta^2}(6 \sin^2 \theta + 4\sin \theta -1) T^2 \!+ \! o(T^3).
	\end{align}
	It is now obvious that $F_Q$ can surpass the Heisenberg limit, $4T^2$, in the small time length region, and that the improvement of precision increases with the strength of decay (i.e., the noisier the better).

\begin{figure}[htb]
	\centering\includegraphics[width=0.47\textwidth]{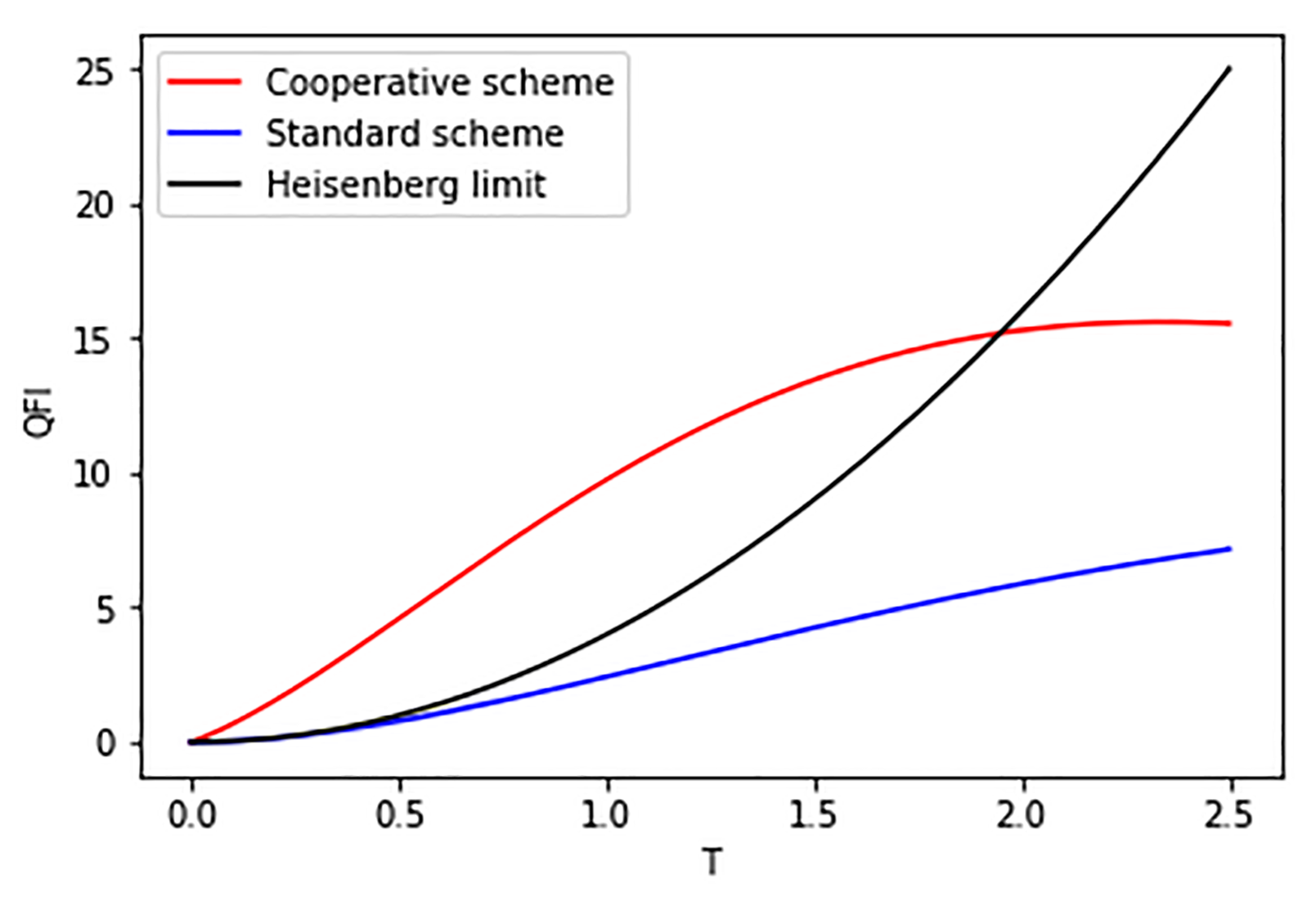}
	\caption{Quantum Fisher information for the single-spin system in the presence of spontaneous emission, under the cooperative  and the standard  schemes respectively, where $B_z = 0.1, B_x = 0.1,\gamma = 0.5$. The Heisenberg limit  under the unitary dynamics is also plotted for comparison.}
	\label{decay_short}
\end{figure}

In Fig. \ref{decay_short} we plot the values of quantum Fisher information $F_Q(\rho_{T})$ as $T$ grows under the cooperative scheme and the standard scheme without coherent control, respectively. We also compare them with the Heisenberg limit under the unitary dynamics.
This is consistent with the quantum Fisher information for the ground state, as in the limit of $T \to \infty$. Namely,
\begin{align}
	\lim_{T \to \infty} F_Q(\rho_{T})  =& \frac{1}{\Delta^2}\left[-\frac{1}{2}(2+\sin{\theta})\sin^3 \theta-\frac{1}{4}\sin^2 \theta (-4 \sin \theta + \cos 2\theta -5) \right]\nonumber\\
	& = \frac{\sin^2 \theta}{\Delta^2} = \frac{B_x^2}{(B_x^2+B_z^2)^2},
\end{align}
which is exactly the quantum Fisher information for the ground state $|g\rangle$.

In addition, from Fig. \ref{decay_short} one can see that the cooperative scheme beats the standard scheme, and the precision limit in the cooperative scheme can even surpass the Heisenberg limit in the region of small $T$ ($T < 2.0$). Please note that this region is where quantum metrology is mostly applied in practical experiments in the presence of noises and phase ambiguity.

\begin{remark}
	The decay rate $\gamma$ can also be made to rely on the parameter to be estimated. For example, if the environment is in a thermal state at the temperature $T_e$, the dynamic equation for the singe-spin evolution is then given by
	\begin{equation}
		\begin{aligned}
			\dot{\rho}= &-i [B_z\sigma_z+B_x\sigma_x,\rho]+\gamma_{+}^H(\sigma_{+}^H \rho \sigma_{-}^H - \frac{1}{2} \{ \sigma_{-}^H \sigma_{+}^H , \rho\})  +\gamma_{-}^H (\sigma_{-}^H \rho \sigma_{+}^H - \frac{1}{2} \{ \sigma_{+}^H \sigma_{-}^H,\rho \}),
		\end{aligned}
	\end{equation}
	where $\gamma_{+}^H=\gamma_0 N$ and $\gamma_{-}^H=\gamma_0(N+1)$ with $N = \frac{1}{e^{\omega/T_e}-1}$ ($\hbar$ is set to be 1). Here $\gamma_0= \frac{4 \omega^3 |\vec d|^2}{3}$ with $\vec d$ the dipole vector, and $\omega=2\sqrt{B_z^2+B_x^2} = 2\Delta$ is the energy gap between the excited state and the ground state \cite{breuer2002theory}. In this case the decay rate $\gamma_+^H$($\gamma_-^H$) also encodes the parameter which provides additional contribution to the parametrization. Fig. \ref{single_para} demonstrates this cooperative effect by plotting the values of QFI in comparison to the Heisenberg limit under the unitary dynamics. 
	\begin{figure}[htb]
		\centering\includegraphics[width=0.53\textwidth]{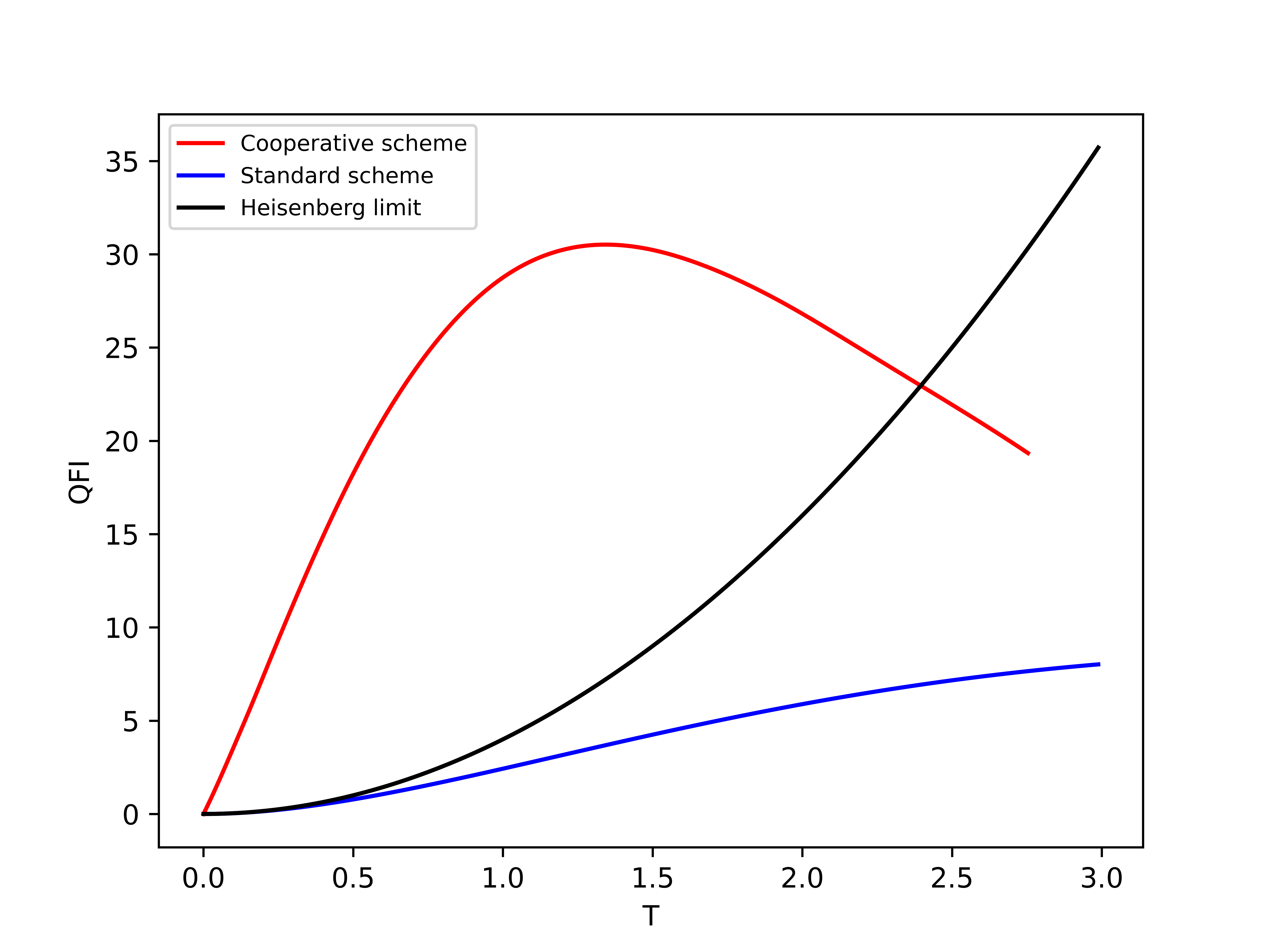}
		\caption{Quantum Fisher information under the cooperative scheme with the environment in the thermal state, where $B_z = 0.3, B_x = 0.1, |\vec d|= 2, T_e = 0$. The values of QFI under the standard scheme, and the Heisenberg limit under the unitary dynamics are also plotted for comparison.}
		\label{single_para}
	\end{figure}
\end{remark}

\subsection{Dephasing noise}

\begin{figure}[htb]
		\centering\includegraphics[width=0.47\textwidth]{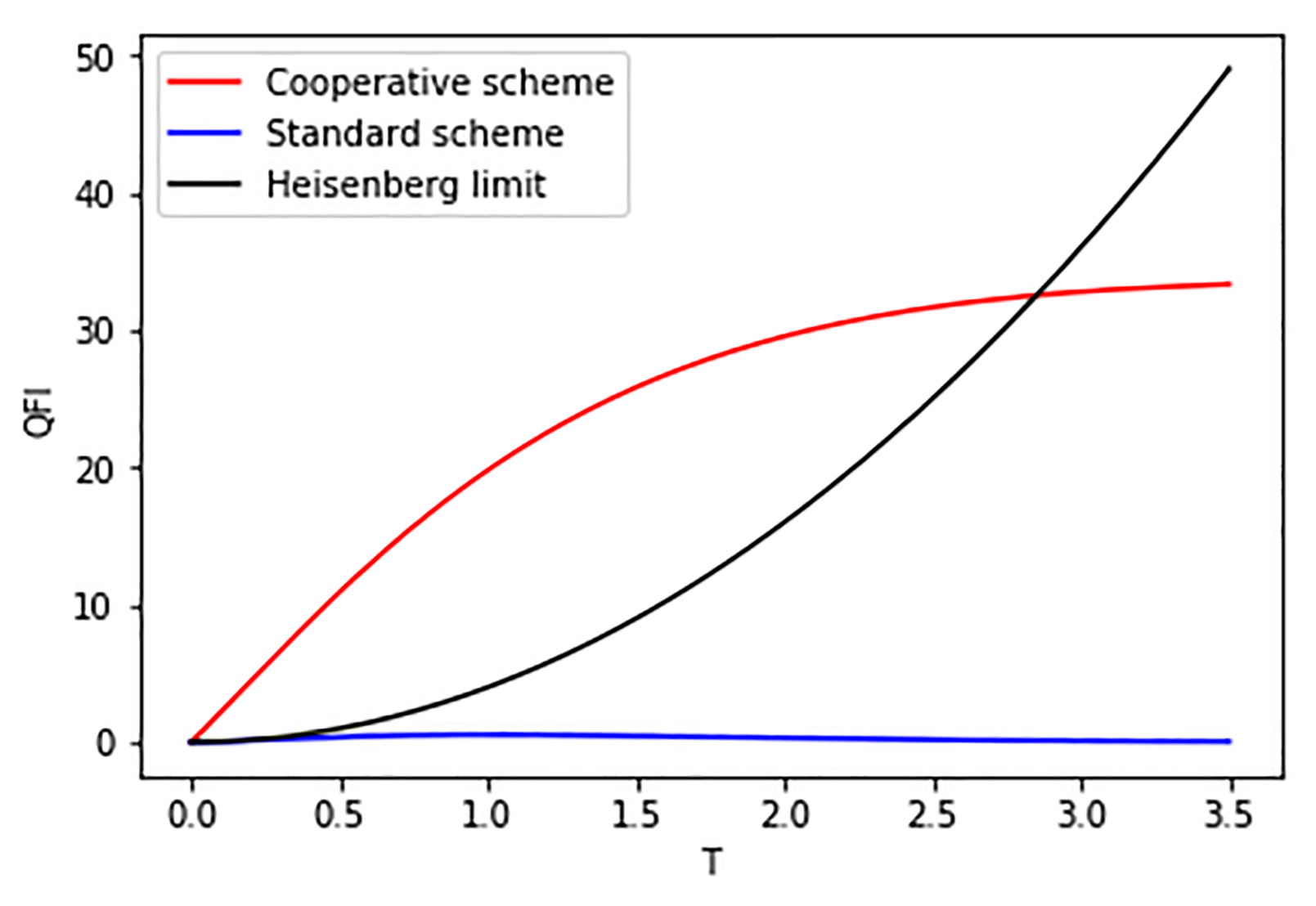}
		\caption{Quantum Fisher information for the system with the dephasing noise, under the cooperative and the standard schemes respectively, where $B_z = 0.1, B_x = 0.1,\gamma= 0.25$. The Heisenberg limit under the unitary dynamics is also plotted for comparison. }
		\label{dephasing_short}
	\end{figure}

It is known that in the presence of dephasing noise, the maximal quantum Fisher information that can be achieved at the time $T$ is $F_Q=4e^{-4\gamma T}T^2$ under standard schemes \cite{yuan2015quantum}, which is also always smaller than the standard Heisenberg limit $4T^2$. However, if we add a controlled field along the $X$-direction, the dynamics for the evolving single qubit satisfy
	\begin{eqnarray}
		\label{eq:controldephasing}
		\dot{\rho} = -i[B_z\sigma_z+B_x\sigma_x,\rho]+\gamma(\sigma_{n}\rho\sigma_{n}-\rho),
	\end{eqnarray}
	where $\sigma_n=\frac{B_z\sigma_z+B_x\sigma_x}{\sqrt{B_z^2+B_x^2}}$ is defined along the direction of the combined field. This is due to the change of axis regarding the spin along the combined field. In order to see the effect of the interplay between coherent control and noise, we first prepare the probe state at $\frac{|0\rangle+|1\rangle}{\sqrt{2}}$ in the original basis. And then we add the controlled field and let the spin evolve according to the dynamics governed by Eq. (\ref{eq:controldephasing}) for the time duration $T$. The QFI $F_Q(\rho_{T})$ can be obtained in the way similar to the scenario of spontaneous emission, having the value greater than $4T^2$ for small $T$.  The effect of the cooperative interaction in this case can be observed from  Fig. \ref{dephasing_short}, where the cooperative scheme not only achieves higher precision limit than the standard scheme does, but also beats the Heisenberg limit, in the time region where practical quantum metrology is mostly used when noises are present. 

\section{Enhanced precision of parameter estimation for multi-spin systems \\under the cooperative scheme}
\label{sec:TSD}

For multiple spins, it has been shown that under the unitary dynamics coupling between different spins cannot contribute to the improvement of precision limit \cite{giovannetti2006quantum,boixo2007generalized}. While by means of the interplay between coherent control and noises, the coupling can help shape the ground and excited states, making them potentially more sensitive to the parameter. In particular, with such coupling, the ground and excited states can be entangled, which thus induces non-local dynamics that could be taken advantage of to improve the precision limit. This can be explained by considering a two-spin system having following Hamiltonian 
\begin{equation}\label{hamE}
	H(B_z)=B_z(\sigma^{1}_{z}+\sigma^{2}_{z})+\sigma^{1}_{z}\sigma^{2}_{z}.
\end{equation}
Here the matrices $\sigma_z^1=\sigma_z\otimes I_2$, and $\sigma_z^2=I_2\otimes \sigma_z$, with $I_2$ denoting the $2\times 2$ identity matrix. Additionally, the Hamiltonian is written in the manner that the coupling strength between these two spins is scaled to be $1$.

Under the unitary dynamics, the highest precision limit is achieved by preparing the probe state at $\frac{|00\rangle+|11\rangle}{\sqrt{2}}$, which gives the maximal quantum Fisher information $F_Q=16T^2$ at the time $T$. In this specific case the coupling does not help improve the precision limit, since the same precision can be achieved without the coupling. Namely, when the Hamiltonian is simply $B_{z}(\sigma^{1}_{z}+\sigma^{2}_{z})$, the maximal quantum Fisher information can also reach $16T^2$ by preparing the probe state at $\frac{|00\rangle+|11\rangle}{\sqrt{2}}$ \cite{giovannetti2006quantum,boixo2007generalized,yuan2015optimal}.

\begin{figure}[htb]
	\centering\includegraphics[width=0.34\textwidth]{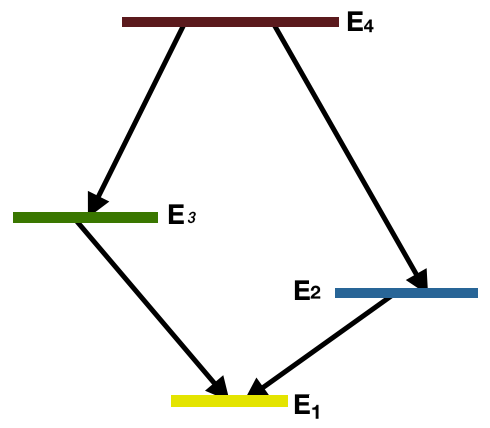}
	\caption{Decay channels induced by a cool reservoir.}
	\label{energy_level}
\end{figure}

We now consider adding a controlled field with $B_x\ll 1$ along the transverse direction, i.e.,
\begin{equation}\label{eq:hamE2q}
	H(B_z)=\sigma^{1}_{z}\sigma^{2}_{z}
	+B_{z}(\sigma^{1}_{z}+\sigma^{2}_{z})+B_x(\sigma_x^1+\sigma_x^2).
\end{equation}
We assume that the system is interacting with a cool reservoir which induces decays between the eigen-states of the system as shown in Fig. \ref{energy_level}, where $E_k$ ($k\in\{1,2,3,4\}$) denote eigen-energies of $H(B_z)$ in Eq.~\eqref{eq:hamE2q} and $E_1<E_2<E_3<E_4$. In the following we demonstrate the utilization of cooperative interaction between coherent control and noises in a two-spin system. 

	Consider the two-spin system described by Eq.~\eqref{eq:bothmaster} by adding a controlled field along the transverse direction, with the Hamiltonian
	\begin{equation*}
		H(B_z)=\sigma^{1}_{z}\sigma^{2}_{z}
		+B_{z}(\sigma^{1}_{z}+\sigma^{2}_{z})+B_x(\sigma_x^1+\sigma_x^2),
	\end{equation*}
	and the Lindblad super-operator
	\begin{align*}
		\mathcal{L}(\rho)=\sum_{(j,k)\in S}\gamma_{jk}\big(| E_k \rangle \langle E_j | \rho | E_j \rangle \langle E_k | - \frac{1}{2}\{ | E_j \rangle \langle E_j | ,\rho\}\big).
	\end{align*}
	Here $S = \{(4,3),(4,2),(3,2),(3,1)\}$, and the associated decay rates $\gamma_{ij} = \frac{4\omega_{jk}^3 |\vec d|^2}{3}$ ($\vec d$ is the dipole vector) are induced by the cool reservoir as depicted in Fig. \ref{energy_level} with $\omega_{jk}=E_j-E_k$.
	
	To view the performance of the cooperative scheme, we first prepare the initial state at $\frac{|00\rangle  + |11 \rangle}{\sqrt{2}}$. And then we add a controlled field along the transverse direction and let the system evolve according to the Hamiltonian given in Eq.~\eqref{eq:hamE2q} with $B_x\ll 1$. 
	
\begin{figure}[htb]
		\centering\includegraphics[width=0.47\textwidth]{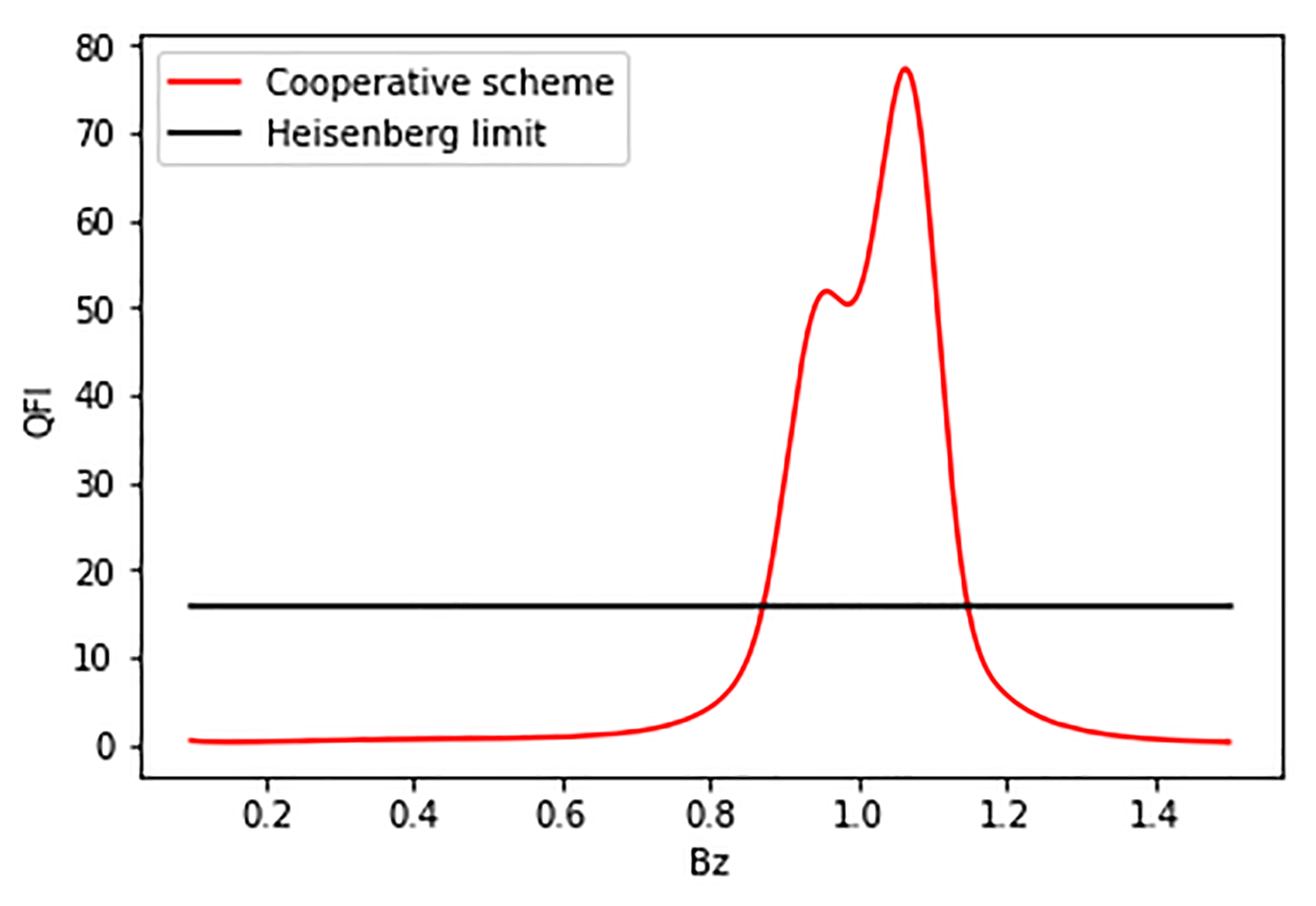}
		\caption{Quantum Fisher information of the cooperative scheme and the Heisenberg limit for different $B_z$, where $B_x = 0.1, T = 1,|\vec d| =10$.}
		\label{double_Bz}
	\end{figure}

	One can write the effective Hamiltonian based on the two lowest energy levels as follows\cite{zhang2008detection}:
	\begin{equation}
		H_{\textrm{eff}} = - |B_z| I + (1 - |B_z|) \sigma_z + \sqrt{2} B_x\sigma_x,
	\end{equation}
	where $I$ denotes the identity operator.
	The two lowest energy states of this effective Hamiltonian are\\
	$|g \rangle = -\sin{\frac{\phi}{2}}|\!\!\uparrow \rangle + \cos{\frac{\phi}{2}}|\!\!\downarrow \rangle$, and  $| e \rangle = \cos{\frac{\phi}{2}} |\!\!\uparrow \rangle + \sin{\frac{\phi}{2}}|\!\!\downarrow \rangle$.  Here the states $|\!\!\uparrow\rangle=|11\rangle$, $|\!\!\downarrow\rangle = \frac{|01\rangle + | 10 \rangle}{\sqrt{2}}$, and  $\tan{\phi} = \frac{\sqrt{2}B_x}{1-B_z}$ for $\phi \in [0, \pi]$. These states can be taken as a good approximation of the two lowest energy states of the actual full Hamiltonian. One can obtain that the quantum Fisher information this approximate ground state $|g \rangle$  is  $\frac{2B_x^2}{(2B_x^2+(-1+B_z)^2)^2}$, which achieves the maximal value at $B_z=1$. 
In Fig. \ref{double_Bz}, we plot the quantum Fisher information at a fixed given time $T =1$ for different values of $B_z$. It can be seen that the quantum Fisher information surpasses the highest precision limit under the ideal unitary dynamics when $B_z \in [0.89,1.14]$. When $B_z$ is outside this region, one can always shift it into this region by means of adaptive control. In the asymptotic limit, the precision can hit the peak at $B_z\approx 1$. In fact, the point $B_z=1$ is a critical one, around which the ground state is the most sensitive to the parameter \cite{zhang2008detection}, which is consistent with our calculation. 
	
\begin{remark}
	Please note that, by tuning the control, we can adjust the highest QFI of the ground state (which is denoted by $F_{\max}=\max_{B_z}F_Q(|g\rangle)$) as well as the width of the region around $B_z\approx 1$ that has the QFI surpassing the Heisenberg limit $16T^2$ (which is denoted by $W=$ the length of the region $\Xi_W = \{B_z | F_Q(|g\rangle) \ge 16T^2\}$). Based on the QFI for the ground state, we can obtain the following relation between $F_{\max}$ and $W$
	\begin{eqnarray}
		\label{eq: tradeoff}
		\frac{W^2}{4} = \frac{1}{\sqrt{F_{\max}}}(\frac{1}{4T}-\frac{1}{\sqrt{F_{\max}}}), \forall F_{\max} >16T^2.
	\end{eqnarray}
	This indicates a trade-off between $F_{\max}$ and $W$, which can be adaptively tuned by the control. In practice, the unknown region for the value of the parameter may be relatively wide at the beginning, we can then use a strong control field, which corresponds to a small value of $F_{\max}$ with a big value of $W$, to accommodate the relatively wide region; after collecting certain amount of measurement data which narrows down the unknown region, we can then adaptively weaken the control field so as to obtain bigger value of $F_{\max}$. Please note that under the standard scheme the term contains the parameter, which is $B_z(\sigma_z^1+\sigma^2_z)$, acts on the probe state locally, whose precision is thus bounded by the Heisenberg limit \cite{yuanfd}. However, under the cooperative scheme, the ground state becomes entangled, which, through the interplay between coherent control and decay, induces non-local dynamics. Or rather say, the cooperative scheme turns local parametrization under the standard scheme to non-local parametrization, which then provides possibilities to go beyond the precision limit under the standard scheme. 
\end{remark}

\begin{remark}
For the two-spin system under the cooperative scheme, as demonstrated in Fig. 6, at $T=1$ the maximal quantum Fisher information that can be achieved is approximately 77. Under the unitary dynamics, the maximal quantum Fisher information is 16 at $T=1$ ($16T^2$ at the time $T$). Therefore, in our scheme the precision is improved by 77-16 = 59 for the two-spin system.
	
	By contrast, for the single-spin system under the cooperative scheme, as demonstrated in Fig. 3, at $T=1$ the maximal quantum Fisher information that can be achieved is approximately 27.5. In comparison to the unitary dynamics under where the maximal quantum Fisher information is $4T^2$ at the time $T$, the precision of the cooperative scheme is improved by 27.5-4 = 23.5 for the single-spin system. 
	
	In view of the above analysis, it can be seen that the improvement of precision is more significant in the multi-spin case.
\end{remark}

\section{Conclusions and future work}
\label{sec:Con}
We first consider the singe-qubit system as given in Eq.~\eqref{eq:bothmaster} in this paper.
It has been shown that with properly designed control, it is possible to encode the unknown parameter $B_z$, which only appears in the Hamiltonian in the previous work, now in both the coherent and noisy parts of the evolution. This enables us to achieve much higher precision limit in comparison with the standard scheme, as stated in Section 2. More precisely, the focus of the cooperative scheme is on the adoption of coherent control to better encode the parameter into multiple components of the dynamics, which thus opens up novel directions for studies of quantum metrology.  It is worth noting that, when designing the control signal, the form of noises needs to be taken into account such that the parameter to be estimated can be encoded in the noisy operator after adding coherent control to the Hamiltonian directly. We are also concerned with multi-spin systems.
It has been shown that for multiple spins, unlike the standard scheme, the coupling between different qubits can help shape the ground and excited states, making them more sensitive to the parameter to be estimated under the cooperative scheme. The ground state can be made entangled due to the interplay between coherent control and noises, which gives non-local parametrization leading to the improvement of precision. These results have been illustrated by the figures and rigorously proved in Section 3. 
In a word, under  cooperative schemes, various factors of the dynamics, such as the couplings and noises, play active roles, as opposed to the passive roles they play under standard schemes.  The  cooperative scheme can be readily implemented on practical systems such as NV-center and superconducting systems \cite{CMZ2018,BSB2020,WWM2019,WCL2022}. While the conventional wisdom to improve the precision is to suppress the noises, the cooperative scheme provides an alternative approach that makes use of the noises which complements the conventional techniques.

 We expect the results of this paper to lead to thorough and systematic exploration of cooperative schemes, together with the associated new ultimate precision limits that go beyond standard schemes.




\medskip
\section*{Acknowledgements}  
The authors acknowledge support by National Natural Science Foundation of China under Grant Nos. 62003113 and 62173296, and by the Research Grants Council of Hong Kong under GRF No. 14307420.\\
Y. Chen and Z. Miao contributed equally to this work.

\section*{Conflict of Interest}
The authors declare no conflict of interest.
\medskip

%

\newpage

\begin{thebibliography}{68}


	\bibitem{giovannetti2011advances}
	 V. Giovannetti, S. Lloyd, L. Maccone,  
	\newblock{\em Nat. Photon.} {\bf 2011}, {\em 5},~222--229.
	

	\bibitem{giovannetti2006quantum}
	 V. Giovannetti, S. Lloyd, L. Maccone,  
	\newblock {\em Phys. Rev. Lett.} {\bf 2006}, {\em 96},~010401.


	\bibitem{anisimov2010quantum}
	P. M. Anisimov, G. M. Raterman, A. Chiruvelli, W. N. Plick, S. D. Huver, H. Lee,  J. P. Dowling,
	\newblock {\em Phys. Rev. Lett.} {\bf 2010}, {\em 104},~103602.

	
	\bibitem{escher2012general}
	 B. M. Escher, R. L. de Matos Filho, L .Davidovich, 
	\newblock {\em Nat. Phys.} {\bf 2011}, {\em 7},~406--411.

	
	\bibitem{joo2011quantum}
	J. Joo, W. J. Munro, T. P. Spiller.
	\newblock {\em Phys. Rev. Lett.} {\bf 2006}, {\em 96},~010401.
	
	
	\bibitem{higgins2007entanglement}
	B. L. Higgins, D. W. Berry, S. D. Bartlett, H. M. Wiseman, G. J. Pryde,
	\newblock {\em Nature} {\bf 2007}, {\em 450},~393--396.

	
	\bibitem{braunstein1996generalized}
	S. L. Braunstein, C. M. Caves, G. J. Milburn,
	\newblock {\em Ann. Phys.} {\bf 1996}, {\em 247},~135--173.
	
	
	\bibitem{demkowicz2014usin}
	R. Demkowicz-Dobrzański, L. Maccone,
	\newblock {\em Phys. Rev. Lett.} {\bf 2014}, {\em 113},~250801.
	

	\bibitem{demkowicz2012elusive}
	R. Demkowicz-Dobrzański, J. Kołodyński, M. Guţă,
	\newblock {\em Nat. Commun.} {\bf 2012}, {\em 3},~1--8.

	
	\bibitem{schnabel2010quantum}
	R. Schnabel, N. Mavalvala, D E. McClelland, P. K. Lam, 
	\newblock {\em Nat. Commun.} {\bf 2010}, {\em 1},~1--10.
	
	
	\bibitem{ligo2011gravitational}
	LIGO,
	\newblock {\em Nat. Phys.} {\bf 2011}, {\em 7},~962--965.
	

	
	\bibitem{lugiato2002quantum}
	L. A. Lugiato, A. Gatti, E. Brambilla,
	\newblock {\em J. Opt. B: Quantum Semiclass. Opt.} {\bf 2002}, {\em 4},~S176.
	

	\bibitem{morris2015imaging}
	P. A. Morris, R. S. Aspden, J. E. Bell, R. W. Boyd, M. J. Padgett, 
	\newblock {\em Nat. Commun.} {\bf 2015}, {\em 6},~1--6.

	
	\bibitem{roga2016security}
	W. Roga, J. Jeffers,
	\newblock {\em Phys. Rev. A} {\bf 2016}, {\em 94},~032301.

	
	\bibitem{tsang2016quantum}
	M. Tsang, R. Nair, Lu X M, 
	\newblock {\em Phys. Rev. X} {\bf 2016}, {\em 6},~031033.
	

	
	\bibitem{buvzek1999optimal}
	V. Bužek, R. Derka, S. Massar,
	\newblock {\em Phys. Rev. Lett.} {\bf 1999}, {\em 82},~2207.
	
	
	\bibitem{leibfried2004toward}
	D. Leibfried, M D. Barrett, T. Schaetz, J. Britton, J. Chiaverini, W. M. Itano, J. D. Jost, C. Langer, D. J. Wineland, 
	\newblock {\em Science} {\bf 2004}, {\em 304},~1476--1478.
	

	
	\bibitem{roos2007designer}
	C F. Roos, M. Chwalla, K. Kim, M. Riebe, R. Blatt, 
	\newblock {\em Nature} {\bf 2006}, {\em 443},~316--319.
	

	\bibitem{borregaard2013near}
	J. Borregaard, A. S. Sørensen,
	\newblock {\em Phys. Rev. Lett.} {\bf 2013}, {\em 111},~090801.
	

	
	\bibitem{ludlow2015optical}
	A. D. Ludlow, M. M. Boyd, J. Ye, E. Peik,  P. O. Schmidt,
	\newblock {\em Rev. Mod. Phys.} {\bf 2015}, {\em 87},~637.
	

	\bibitem{shapiro2009quantum}
	J. H. Shapiro, S. Lloyd,
	\newblock {\em New. J. Phys.} {\bf 2009}, {\em 11},~063045.
	

	
	\bibitem{lopaeva2013experimental}
	E. Lopaeva, I. R. Berchera, I. Degiovanni, S. Olivares, G. Brida, M.Genovese,
	\newblock {\em Phys. Rev. Lett.} {\bf 2013}, {\em 110},~153603.
	

	
	\bibitem{dowling1998correlated}
	J. P. Dowling,
	\newblock {\em Phys. Rev. A} {\bf 1998}, {\em 57},~4736.

	
	\bibitem{JCJD2021}
	J. P. Cooling, J. A. Dunningham,
	\newblock {\em  J. Phys. B: At. Mol. Opt. Phys.} {\bf 2021}, {\em 54},~195502.

	
	\bibitem{huelga1997improvement}
	S. F. Huelga, C. Macchiavello, T. Pellizzari,  A. K. Ekert, M. B. Plenio, J. I. Cirac, 
	\newblock {\em Phys. Rev. Lett.} {\bf 1997}, {\em 79},~3865.
	

	
	\bibitem{chin2012quantum}
	A. W. Chin, S. F. Huelga, M. B. Plenio,
	\newblock {\em Phys. Rev. Lett.} {\bf 2012}, {\em 110},~233601.
	

	
	\bibitem{HallPRX}
	M. J. W. Hall, H. M. Wiseman,
	\newblock {\em Phys. Rev. X} {\bf 2012}, {\em 2},~041006.
	
	
	\bibitem{Berry2015}
	D. W. Berry, M. Tsang, M. J. W. Hall, H. M.Wiseman,
	\newblock {\em Phys. Rev. X} {\bf 2015}, {\em 5},~031018.
	
	
	%
	
	\bibitem{Alipour2014}
	S. Alipour, M. Mehboudi, A. T. Rezakhani,
	\newblock {\em Phys. Rev. Lett.} {\bf 2014}, {\em 112},~120405.
	

	\bibitem{Beau2017}
	M. Beau, A. del Campo,
	\newblock {\em Phys. Rev. Lett.} {\bf 2017}, {\em 119},~010403.
		

	
	\bibitem{Holevo}
	A. Holevo,
	\newblock {\em Probabilistic and Statistical Aspects of Quantum Theory},
	North-Holland, Amsterdam {\bf 1982}.
	

	
	\bibitem{helstrom1976quantum}
	C. W. Helstrom,
	\newblock {\em Quantum Detection and Estimation Theory}, 
	\newblock Academic Press, New York {\bf 1976}.
	

	
	\bibitem{boixo2007generalized}
	S. Boixo, S. T. Flammia, C. M. Caves, J. M. Geremia,
	\newblock {\em Phys. Rev. Lett.} {\bf 2007}, {\em 98},~090401.
	

	
	\bibitem{yuan2015optimal}
	H. Yuan, C.-H. F. Fung,
	\newblock {\em Phys. Rev. Lett.} {\bf 2015}, {\em 115},~110401.
	

	
	\bibitem{yuan2016sequential}
	H. Yuan,
	\newblock {\em Phys. Rev. Lett.} {\bf 2016}, {\em 117},~160801.

	
	
%
%
%
%
%
%
	

	

	
	

	
	\bibitem{Dur2014}
	W. D\"ur, M. Skotiniotis, F. Fr\"owis, B. Kraus,
	\newblock {\em Phys. Rev. Lett.} {\bf 2014}, {\em 112},~080801.

	
	\bibitem{Arrad2014}
	G. Arrad, Y. Vinkler, D. Aharonov, A. Retzker, 
	\newblock {\em Phys. Rev. Lett.} {\bf 2014}, {\em 112},~150801.

	
	\bibitem{Kessler2014}
	E. M. Kessler, I. Lovchinsky, A. O. Sushkov, M. D. Lukin,
	\newblock {\em Phys. Rev. Lett.} {\bf 2014}, {\em 112},~150802.
	

	
	\bibitem{Unden2016}
	T. Unden, P. Balasubramanian, D. Louzon, Y. Vinkler, M. B. Plenio,
	\newblock {\em Phys. Rev. Lett.} {\bf 2016}, {\em 116},~230502.

	
	
	\bibitem{Zhou2018}
	S. Zhou, M. Zhang, J. Preskill, L. Jiang,
	\newblock {\em Nat. Commun.} {\bf 2018}, {\em 9},~1--11.

	
	\bibitem{SMMN2021}
	N. Shettell, W. J. Munro, D. Markham, K. Nemoto,
	\newblock {\em New J. Phys.} {\bf 2021}, {\em 23},~043038.
	
	
	
	\bibitem{Schmitt832}
	S. Schmitt, T. Gefen, F. M. Stürner, T. Unden, G. Wolff,
	\newblock {\em Science} {\bf 2017}, {\em 356},~832--837.

	

	\bibitem{Boss837}
	J. M. Boss, K. S. Cujia, J. Zopes, L. C. Degen,
	\newblock {\em Science} {\bf 2017}, {\em 356},~837--840.
	
	

	\bibitem{SekatskoNJP2016}
	P. Sekatski, M. Skotiniotis, W. Dür,
	\newblock {\em New. J. Phys.} {\bf 2016}, {\em 18},~073034.
	

	\bibitem{LangPRX2015}
	J. E. Lang, R. B. Liu, T. S. Monteiro. 
	\newblock {\em Phys. Rev. X} {\bf 2015}, {\em 5},~041016.
	

	
	\bibitem{Taylor2008}
	J. M. Taylor, P. Cappellaro, L. Childress, L. Jiang, D. Budker, P. R. Hemmer, A. Yacoby, R. Walsworth, M. D. Lukin,
	\newblock {\em Nat. Phys.} {\bf 2008}, {\em 4},~810--816.
	

	
	\bibitem{Cooper2014}
	A. Cooper, E. Magesan, H. N. Yum, P. Cappellaro,
	\newblock {\em Nat. Commun.} {\bf 2014}, {\em 5},~1--7.
	

	
	\bibitem{LiuSingle}
	J. Liu, H. Yuan,
	\newblock {\em Phys. Rev. A} {\bf 2017}, {\em 96},~012117.
	

	
	\bibitem{LiuMulti}
	J. Liu, H. Yuan.,
	\newblock {\em Phys. Rev. A} {\bf 2017}, {\em 96},~042114.

	
	\bibitem{LZCWY2022}
	J. Liu, M. Zhang, H. Chen, L. Wang, H. Yuan,
	\newblock {\em Adv. Quantum Technol.} {\bf 2022}, {\em 5},~2100080.
	

	\bibitem{Goldstein2011}
	G. Goldstein, P. Cappellaro, J. R. Maze, J. S. Hodges, L. Jiang, A. S. Srensen, M. D. Lukin,
	\newblock {\em Phys. Rev. Lett.} {\bf 2011}, {\em 106},~140502.
	

	\bibitem{Cappellaro2012}
	P. Cappellaro, G. Goldstein, J. S. Hodges, L. Jiang, J. R. Maze, A. S. SRensen, M. D. Lukin,
	\newblock {\em Phys. Rev. A} {\bf 2012}, {\em 85},~032336.
	
	

	\bibitem{yuan2015quantum}
	H. Yuan, C. H. F. Fung,
	\newblock {\em Phys. Rev. Lett.} {\bf 2015}, {\em 115},~110401.

	
	
	\bibitem{breuer2002theory}
	H. P. Breuer, F. Petruccione,
	\newblock {\em The Theory of Open Quantum Systems,}
	\newblock Oxford University Press, New York {\bf 2022}.
	
	
	\bibitem{zhang2008detection}
	J. Zhang, X. Peng, N. Rajendran, D. Suter,
	\newblock {\em Phys. Rev. Lett.} {\bf 2008}, {\em 100},~100501.

	
	\bibitem{yuanfd}
	H. Yuan, C. H. F. Fung,
	\newblock {\em New. J. Phys.} {\bf 2017}, {\em 19},~113039.

	
	\bibitem{dittmann1999explicit}
	J. Dittmann,
	\newblock {\em J. Phys. A: Math. Gen.} {\bf 1999}, {\em 32},~2663.
	
	\bibitem{CMZ2018}
	M. Cheng, C. Meng, Q. Zhang, C. Duan, F. Shi, J. Du
	\newblock {\em Natl. Sci. Rev.} {\bf 2018}, {\em 5},~346--355.

	
	\bibitem{BSB2020}
	J. F. Barry, J. M. Schloss, E. Bauch, M. J. Turner, C. A. Hart, L. M. Pham, R. L. Walsworth,
	\newblock {\em Rev. Mod. Phys.} {\bf 2020}, {\em 92},~015004.

	
	\bibitem{WWM2019}
	W. Wang, Y. Wu, Y. Ma, W. Cai, L. Hu, X. Mu, Y. Xu, Z. Chen, H. Wang, Y. Song,
	\newblock {\em Nat. Commun.} {\bf 2019}, {\em 10},~4382.
	
	\bibitem{WCL2022}
	W. Wang, Z.-J. Chen, X. Liu, W. Cai, Y. Ma, X. Mu, L. Hu, Y. Xu, H. Wang, Y. P. Song,
	\newblock {\em Nat. Commun.} {\bf 2022}, {\em 13},~3214.

	
\end{thebibliography}

\end{document}